\documentclass[11pt,twoside]{article}

\usepackage{natbib}
\usepackage{asp2004}
\usepackage{epsf}
\usepackage{psfig}
\usepackage{lscape}
\usepackage{graphicx}
\usepackage[hypertex]{hyperref}

\newcommand{\unit}[1]{\mathrm{#1}}

\newcommand{\Halpha}{\ensuremath{\mathrm{H}{\alpha}}}

\newcommand{\sbnn}{{\sc starburst99}}

\begin{document}
\title{A Simple Model For Mid-Infrared
Emission from Normal Galaxies}   %%% Fill in title
\author{B. Nikolic$^{1,2}$, P. Alexander$^{2}$ and D. Ford$^{2}$}   %%% Fill in author names
\affil{$^{1}$NRAO, 520 Edgemont Road, Charlottesville, Va 22903, USA\\
$^{2}$Astrophysics Group, Cavendish Laboratory, Cambridge CB3 0HE, UK}    %%% Fill in author affiliations

\begin{abstract} %%% Abstract to run on from here.
We have combined up-to-date stellar population synthesis models, a
simple radiative transfer approach, and a fully comprehensive dust
model with the aim of developing a simple but quantitative way of
interpreting the mid-infrared spectra of galaxies. We apply these
models to the observed correlations of mid-infrared luminosities (at 8
and 24\,\micron) with the star-formation rate of normal galaxies and
find that the observations are naturally reproduced by our models. We
further find that the observed 24\,\micron\ correlation places a weak
constraint on relative distribution of dust and stars.
\end{abstract}

%%% MAIN BODY OF TEXT GOES HERE. CONSULT "INSTRUCTIONS FOR AUTHORS USING
%%% LATEX2E MARKUP", SECTIONS 2.3-2.6 FOR HELP WITH EQUATIONS, FIGURES,
%%% AND TABLES.

%\section{}   %%% Top level section head (remove "%" symbol)
%\subsection{}   %%% Second level section head (remove "%" symbol)
%\subsubsection{}   %%% Lowest level section head (remove "%" symbol)
%\section*{}	%%% Unnumbered top level section head (remove "%" symbol)
%\subsection*{}   %%% Unnumbered second level section head (remove "%" symbol)

\section{Introduction}

Mid-infrared (mid-IR) emission from galaxies provides information on
their energetics which is complementary to the far-infrared (far-IR)
and arguably more useful. The reason for this is that the very small
dust grains, which produce this emission, can be stochastically heated
\citep{1976ApJ...206..685P} by energetic photons, so that their peak
temperatures are independent of their time-averaged temperatures, in
which case the colour temperature of their emission is independent of
the intensity of the radiation which heats them
\citep{1984ApJ...277..623S}. Consequently, when dust in this regime
dominates the mid-IR luminosity of a galaxy, this luminosity, even
when estimated from a narrow band filter, is an intrinsically weak
function of the spatial distribution of the dust.

This has motivated us to develop a simple model for the infrared
emission from galaxies, concentrating on those wavelengths where
transient heating can dominate. Clearly, this requires an accurate
treatment of transiently heated dust and so we use the
state-of-the-art techniques presented by \cite{2001ApJ...551..807D},
\cite{2001ApJ...554..778L} and \cite{2001ApJ...548..296W}. We make the
connection to galaxy energetics by using \sbnn\
\citep{1999ApJS..123....3L} simple stellar population models, and
simple radiative transfer, to calculate the radiation field which is
heating the dust.

\section{Method}

The modelling procedure we employ consists of three distinct parts:
(1) Calculating the radiation field heating the dust, taking into
account geometric effects and absorption by the dust; (2) Calculating
the infrared radiation emitted by the dust; and, (3) Calculating the
subsequent re-absorption of this infrared radiation by any further
dust it encounters.

As mentioned above, we assume the dust to be heated by a single
population of stars modeled using \sbnn\ \citep{2005ApJ...621..695V}.
Being interested primarily in the transiently heated regime, we
consider a simple configuration in which the dust is in a
geometrically thin spherical shell of radius $R$, with optical depth
set by its total column density of hydrogen atoms $n$. The radiation
field which heats the dust is calculated by assuming it to emanate
from a central source and be isotropic and conserved at radii less
than that of the dust shell.  Absorption of radiation as it traverses
the shell is taken into account although scattering is neglected.  For
consistency, the absorption cross section of the dust is derived from
the same dust model used to calculate the emission from the dust.
Although this model -- a single heating source surrounded by a single
dust shell -- is a poor approximation to most normal star-forming
galaxies, it is entirely equivalent to a more realistic model. If we
alternatively assume star-formation to be distributed over $N$
independent sites, each with an equal star-formation rate, and a dust
shell at the same radius $R$, this is equivalent to putting all of the
star-formation in a single region and the dust at an effective radius
$R_{\rm eff}= R\sqrt{N}$.

The emission from the dust is calculated using the
`thermal-continuous' model of \cite{2001ApJ...551..807D} and the dust
model parameters derived by \cite{2001ApJ...554..778L} and
\cite{2001ApJ...548..296W} for the local interstellar medium. The
final stage of the model is to take account of the re-absorption of
this emission by further dust. Here we make another simplification by
assuming that if the total column density of the dust shell is $n$,
then any emission from dust at a point with column density to heating
source is $n'$ will itself be attenuated by a further column density
$n-n'$ before escaping the shell.

\section{Results}

To investigate the applicability of these models to real galaxies, we
tested how well they reproduce the observed correlation between mid-IR
luminosities and star-formation rates (as traced by
extinction-corrected \Halpha\ luminosities) of galaxies in the {\it
Spitzer} extragalactic First Look Survey (FLS) field
\citep[][]{2005ApJ...632L..79W}. We repeated the analysis of
\cite{2005ApJ...632L..79W} to obtain $K$-corrected 8\,\micron\ and
24\,\micron\ dust luminosities and extinction and aperture corrected
\Halpha\ data for these galaxies.  We exclude galaxies dominated by
active galactic nuclei by using the standard emission-line
diagnostics.  Model mid-IR luminosities were obtained by convolving
our synthetic dust spectra (for a range of star-formation rates) with
the {\it Spitzer\/} filter response curves.  The \Halpha\ luminosities
for the same range of star-formation rates were obtained by converting
the ionising photon production rate from \sbnn\ stellar population
models.

\begin{figure}
  \includegraphics[clip,width=0.49\linewidth]{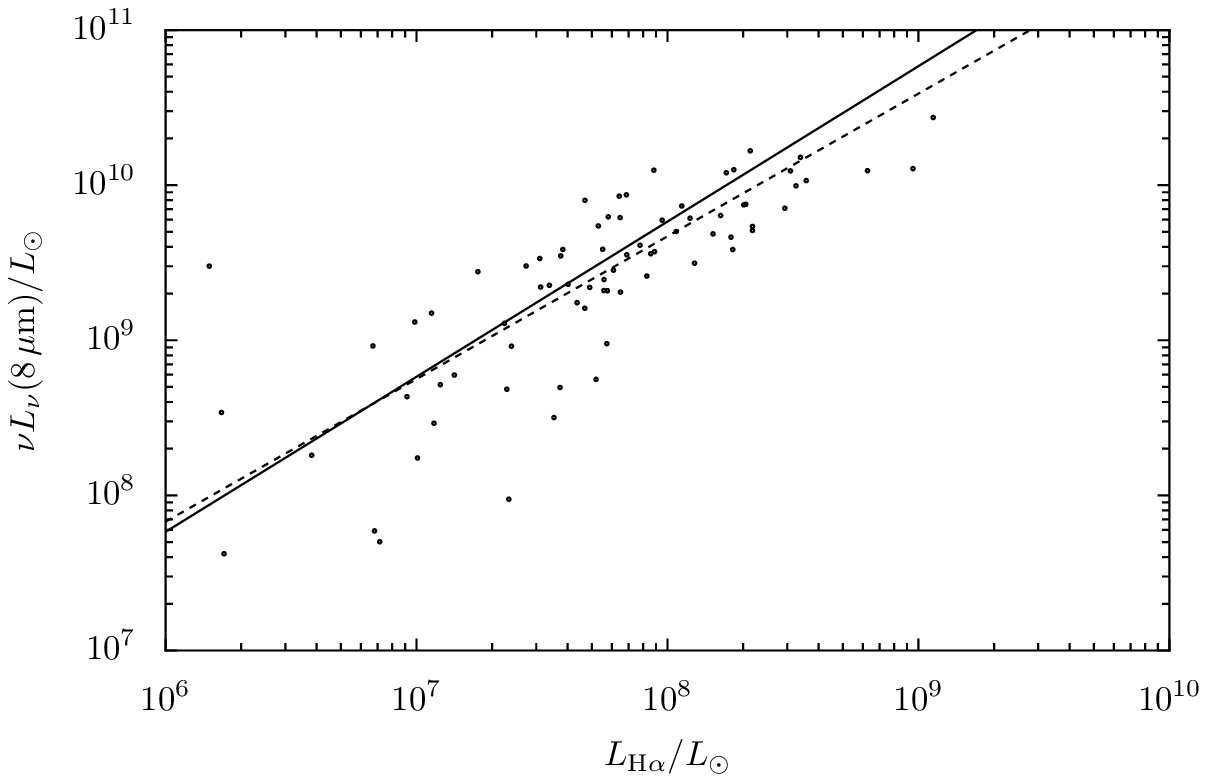}
  \includegraphics[clip,width=0.49\linewidth]{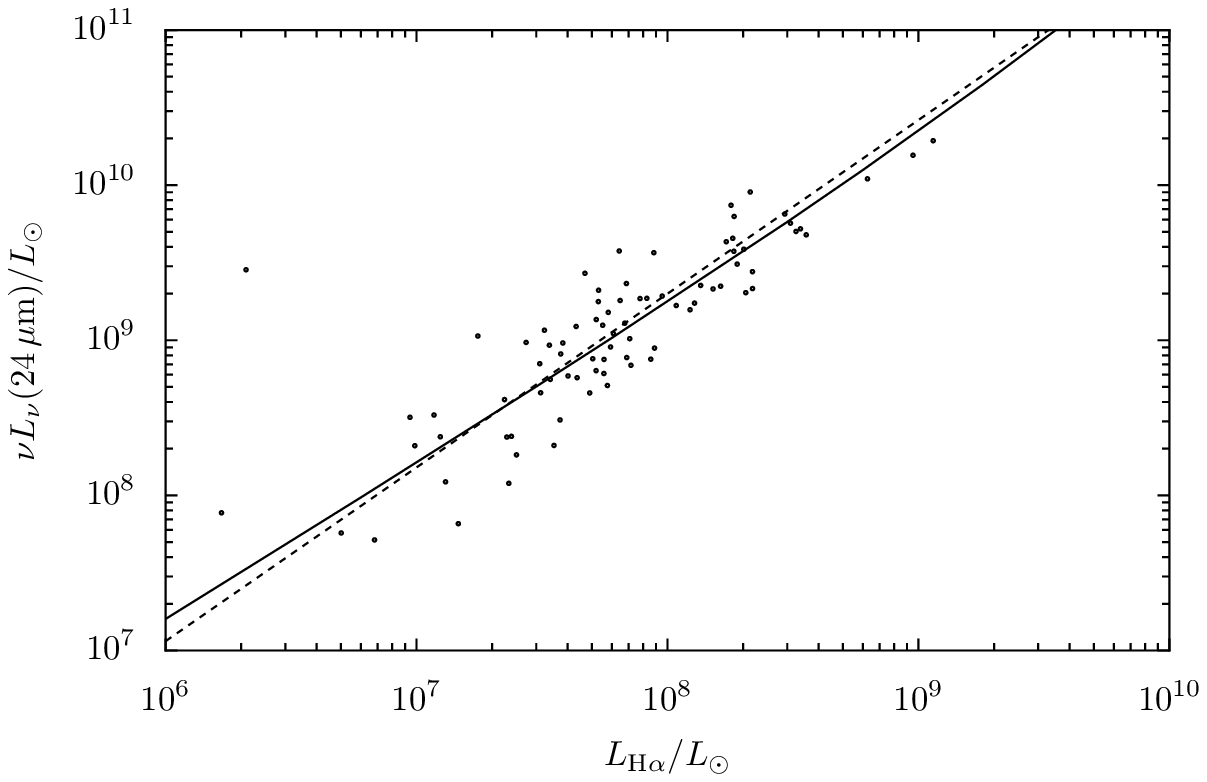}
  \caption{The correlation observed in the {\it Spitzer\/} FLS field
  between mid-IR and \Halpha\ luminosities (points), together with our
  models (solid lines) and the best-fitting power-law model of
  \citet[][dashed lines]{2005ApJ...632L..79W}. The left plot is
  for the IRAC 8\,\micron\ channel; the right for the MIPS
  24\,\micron\ channel.}
  \label{fig:flscorrs}
\end{figure}

\begin{figure}

  \includegraphics[clip,width=0.48\linewidth]{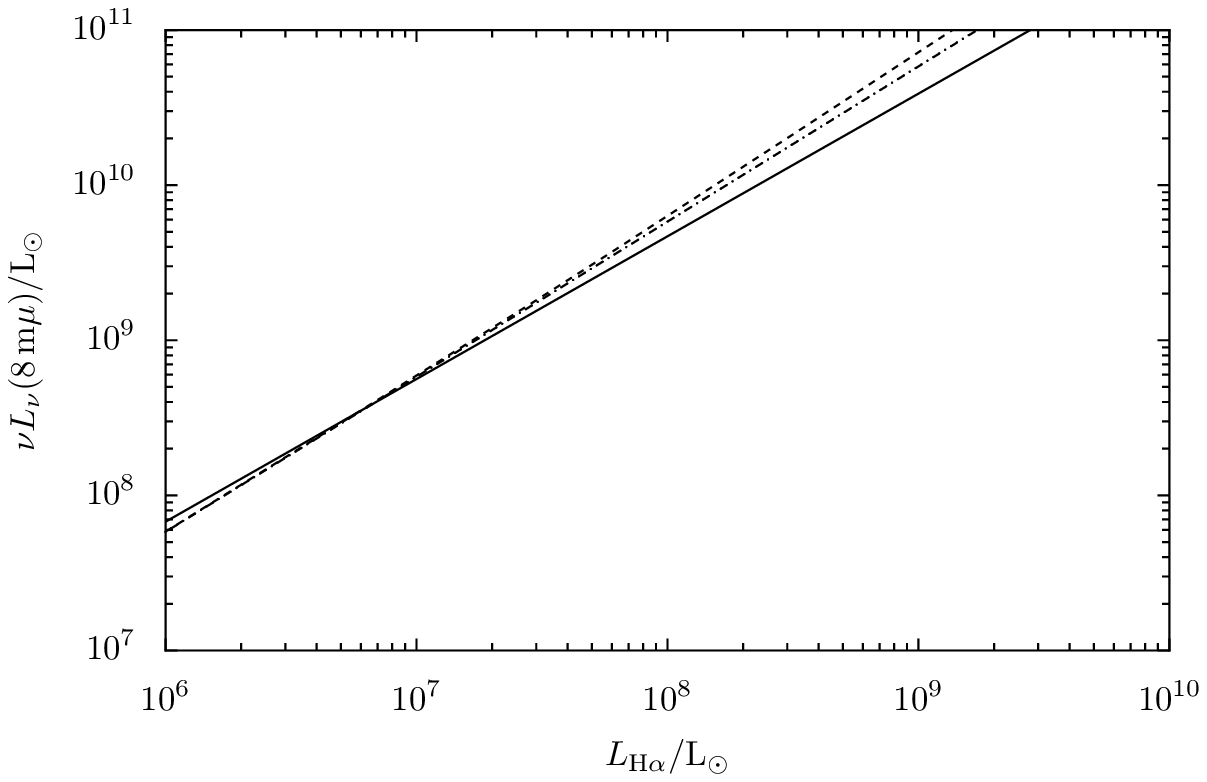}
  \includegraphics[clip,width=0.48\linewidth]{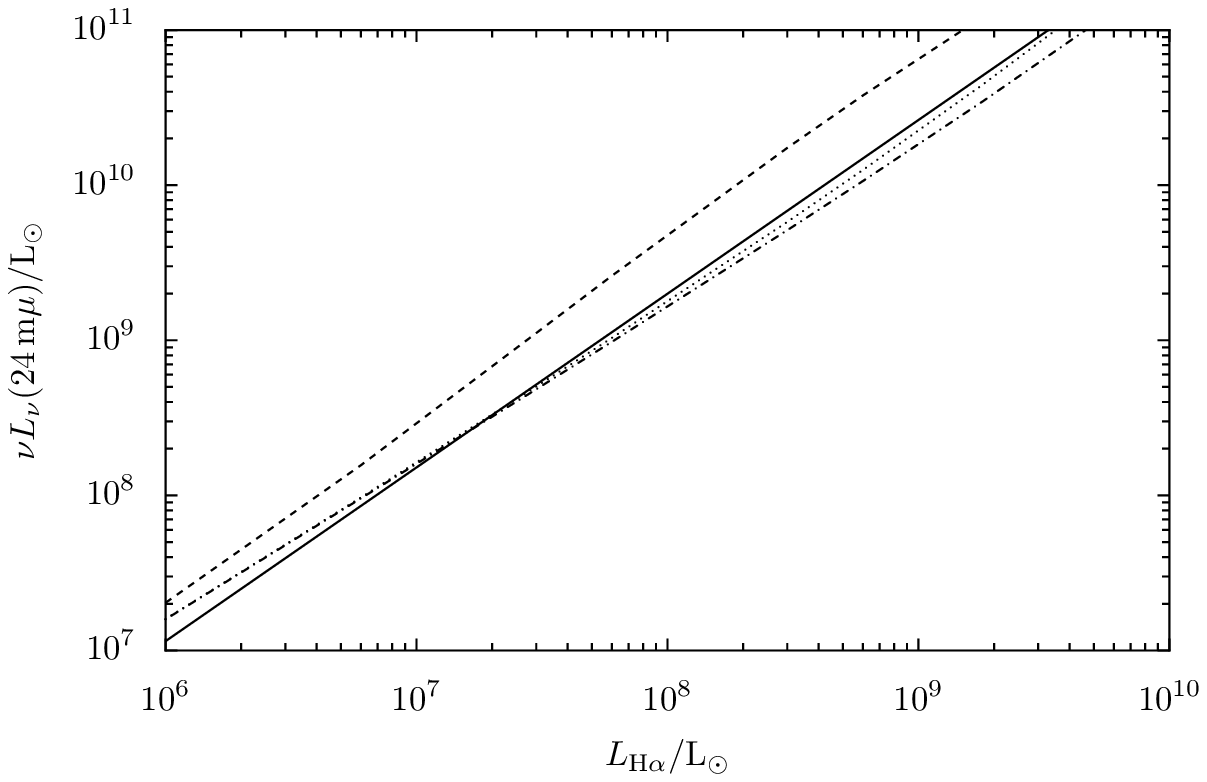}

  \caption{A comparison of the predicted dust
  luminosities (broken line styles) as a function of \Halpha\
  luminosity and the observed best-fitting power-law correlation
  (solid line). The shell radii were 100\,pc, 2\,kpc and 5\,kpc for
  the dashed, dotted and dash-dot-dash line styles respectively (the
  2\,kpc and 5\,kpc traces essentially overlay each other). The left plot is
  for the IRAC 8\,\micron\ channel; the right for the MIPS
  24\,\micron\ channel.}

  \label{fig:WumCmp}

\end{figure}

The results of this comparison are shown in Figure~\ref{fig:flscorrs},
which plots the observed data points on the mid-IR vs \Halpha\
luminosity plane, together with the best-fitting power law curve
published by \citet[][dashed line]{2005ApJ...632L..79W} and the
expected model correlation for $R_{\rm eff}=2\,\unit{kpc}$ (solid
line). With the dust model parameters fixed to Galactic values, the
only free parameter we consider in the models is $R_{\rm eff}$, the
effective radius. We chose $R_{\rm eff}=2\,\unit{kpc}$ for the models
shown in Figure~\ref{fig:flscorrs} as it yielded the best fit to the
observed correlation. The effect that varying $R_{\rm eff}$ has on the
model correlations is shown in the left and right panels of
Figure~\ref{fig:WumCmp} for the 8\,\micron\ and 24\,\micron\ mid-IR
luminosities respectively.

\section{Discussion and Conclusions}

It can be seen from the model correlation between mid-IR and \Halpha\
luminosities, shown by the solid lines in Figure~\ref{fig:flscorrs},
that mid-IR luminosity is largely a linear tracer of the
star-formation rate \emph{if} all of the model parameters are
constant, as would be expected for transiently heated dust. The figure
also shows that the 8\,\micron\ luminosity is in theory more linear
than the 24\,\micron\ luminosity.  This is the case because
multiple-photon heating of grains can be a significant source of
emission at 24\,\micron\ at high star-formation rates.

It is, however, the comparison with real galaxies that is of most
interest. In Figure~\ref{fig:flscorrs}, we also show the observed
positions on the luminosity-luminosity plane of galaxies in the FLS as
well as the best-fitting power-law curve of
\cite{2005ApJ...632L..79W}. Given the relative simplicity of the
models we describe, the agreement with the data is remarkably good. In
particular, we find that the constant of proportionality between
mid-IR and \Halpha\ luminosities is easily reproduced by our models,
which may be interpreted as evidence that transiently heated dust
dominates mid-IR luminosities of galaxies and that most of the
radiation produced by young stars is absorbed by the dust rather than
escape the galaxy. 

There is, however, a significant deviation between the data and the
model for the 8\,\micron\ luminosity. At low \Halpha\ luminosities,
some of the galaxies have much smaller mid-IR luminosities than would
be expected from the model. This has been observed by
\cite{2005ApJ...632L..79W} and interpreted as due the now well-known
deficiency of mid-IR emission from dwarf galaxies
\citep[e.g.,][]{2004ApJS..154..211H}.  There is also a trend for
galaxies with \emph{high} \Halpha\ luminosities to have smaller mid-IR
luminosities then predicted by the model. This is reflected in the
best-fitting power law of \cite{2005ApJ...632L..79W} which has a slope
smaller than one. This deviation is not reproduced by our models which
have a slope close to one.  The reason for the relatively low
8\,\micron\ emission from the most intensely star-forming galaxies is
not clear.  It may due to increasingly high obscuring column
densities, causing the 8\,\micron\ mid-IR emission to become
re-absorbed. Alternatively, it may be due to destruction of the
smallest dust grains, e.g., those with radii around 3.5\,\AA\ to
5\,\AA, which dominate the emission at 8\,\micron.

In contrast, the correlation between the 24\,\micron\ and \Halpha\
luminosities appears to follow our model well at high \Halpha\
luminosities.  The larger size of grains responsible for 24\,\micron\
emission makes it possible to place a constraint on the relative
distribution of dust and the heating sources. Figure~\ref{fig:WumCmp}
shows that effective radii in the range 1--5\,kpc give a good
agreement with the observations. The implication is that the dust
responsible for the bulk of 24\,\micron\ emission is not necessarily
very close to regions of active star formation or very hot. The
observed correlation is well reproduced by placing the dust relatively
far from the heating sources.

\acknowledgements %%% Text of acknowledgements runs on after this
%%% command.

We would like to thank the {\it Spitzer} FLS group for making
available the data and source catalogs at 8\,\micron, C. Papovich for
making available the 24\,\micron\ source catalog
\citep{2005astro.ph.12623P} and the group at MPA for making providing
emission line measurements of SDSS galaxies in the
FLS\footnote{\href{http://www.mpa-garching.mpg.de/SDSS/DR4}{http://www.mpa-garching.mpg.de/SDSS/DR4}}.

%%% THE BIBLIOGRAPHY
%%%
%%% CONSULT SECTION 3 OF "INSTRUCTIONS FOR AUTHORS" FOR HOW TO USE NATBIB.
%%% AUTHORS ARE ENCOURAGED TO USE EITHER THE "THEBIBLIOGRAPY" ENVIRONMENT
%%% BY UNCOMMENTING (DELETING THE "%" SYMBOL) THE COMMANDS BELOW, OR BY
%%% USING THE BIBTEX ENVIRONMENT. TO FIND OUT WHICH IS APPLICABLE TO YOUR
%%% CONTRIBUTION, CONSULT THE VOLUME EDITORS FOR YOUR PROCEEDINGS.
%%%

\end{document}